# The exploitation of Multiple Feature Extraction Techniques for Speaker Identification in Emotional States under Disguised Voices

Noor Ahmad Al Hindawi
*Dept. of Electrical Engineering*
*University of Sharjah*
Sharjah, United Arab Emirates
noorahmedalhindawi@gmail.com

Ismail Shahin
*Dept. of Electrical Engineering*
*University of Sharjah*
Sharjah, United Arab Emirates
ismail@sharjah.ac.ae

Ali Bou Nassif
*Dept. of Computer Engineering*
*University of Sharjah*
Sharjah, United Arab Emirates
anassif@sharjah.ac.ae

*Abstract*— Due to improvements in artificial intelligence, speaker identification (SI) technologies have brought a great direction and are now widely used in a variety of sectors. One of the most important components of SI is feature extraction, which has a substantial impact on the SI process and performance. As a result, numerous feature extraction strategies are thoroughly investigated, contrasted, and analyzed. This article exploits five distinct feature extraction methods for speaker identification in disguised voices under emotional environments. To evaluate this work significantly, three effects are used: high-pitched, low-pitched, and Electronic Voice Conversion (EVC). Experimental results reported that the concatenated Mel-Frequency Cepstral Coefficients (MFCCs), MFCCs-delta, and MFCCs-delta-delta is the best feature extraction method.

## I. INTRODUCTION

Speech is a type of communication that is universal. The method of recognizing the speaker based on the vocal qualities of a given utterance is known as Speaker Recognition (SR) [1]. SR is divided into Speaker Identification (SI) and Speaker Verification (SV). SR is based on finding and extracting distinct aspects of a speaker's speech. Voice biometrics refers to the characteristics of a person's voice. The detailed taxonomy of SR is shown in Figure 1 [2]. SI is the way of identifying the speaker from a given utterance by comparing the utterance's voice biometrics to previously registered utterance models. Various techniques are introduced in the field of SI such as the work by Shahin *et al.* [3]. The SV system, on the other hand, is concerned with recognizing or rejecting the identification of the affirmed speaker.

The "human voice" has a diverse variety of variances due to many extrinsic (such as nearby equipment, reverberation, noise, etc.) and intrinsic elements (such as health psychological status, alcohol, drugs, emotion, etc.) [4]. The deliberate alteration of a person's apparent gender, age, personality, and identity is known as voice disguise [5]. Disguised voices can be detected electronically by altering sound qualities before they reach the listener, or mechanically by interrupting the speech production mechanism in some way [5]. Audio disguise can distort the sound and create a lot of fluctuation in characteristics including pitch, segment duration, formant frequency, and bandwidth [6].

The unidentified speaker would be identified by detecting the specific aspects of the spoofed speeches. The prosody, spectral pattern, and language content are the three main characteristics of the human voice [7]. Low pitched voice, hand over mouth, electronic voice, conversion, high pitched voice, electronic voice transformation, pinched nose voice, and so on are all examples of disguised voices. Voice concealment is used in applications such as speech synthesis, forensic science, and leisure [5].

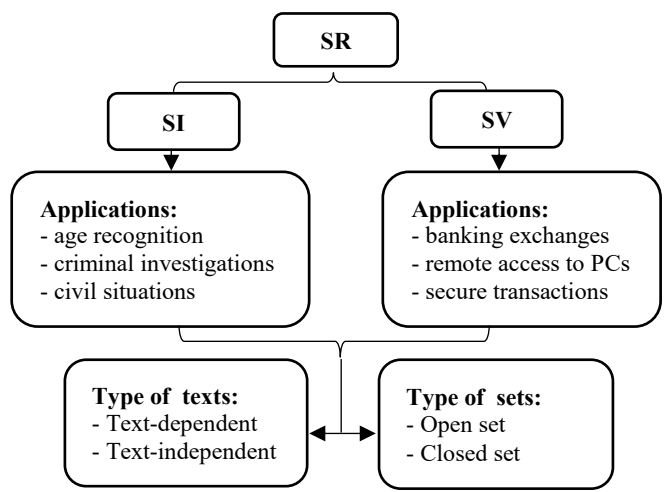

*Figure 1. The SR classification [2]*

The remainder of this work is arranged in the following manner. The background of the five feature extraction approaches is presented in Section II. The Arabic-Emirati speech database is displayed in Section III. In Section IV, the results and comments are presented. Finally, Section V presents the work's conclusion.

## II. BACKGROUND

The feature extraction approaches used are Linear Predictive Coding (LPC) [8], [9], Mel-Frequency Cepstral Coefficients delta-delta (MFCCs $\Delta^2$) [10], Hybrid Algorithm DWPD [11], Probabilistic Linear Discriminate Analysis (PLDA) [12], [13], and Discrete Cosine Transforms (DCT) [14], where the optimal dimensionality result can be achieved. Neither technique is superior to the other; nonetheless, the optimum approach should be determined by the model's performance results. The rationale for using MFCCs $\Delta^2$ method rather than the normal MFCCs, is because MFCCs $\Delta^2$ enhances the performance by 20% due to its aptitude to convey more sufficient information in addition to its ability to attain less computational complexity [15], [16], [17]. Numerous studies have used MFCCs as the optimum feature extraction method for their system [18], [19], [20], [21], [22], [23], [24]. MFCC also recognizes the best output features in the proposed study, where MFCCs were improved in numerous ways to make them quicker, less sensitive to noise, and more stable.

### A. Mel Frequency Cepstral Coefficients (MFCCs)

In speech signals, human perception of sound frequency contents demonstrated by research does not stick to a linear scale. Therefore, each voice with a definite frequency, (f) has a subjective pitch that is calculated on a scale called the 'Mel' scale as given below [25], [14],

$$f_{mel} = 2595 \log_{10}(1 + \frac{f}{700}) \quad (1)$$

where the subjective pitch in 'Mels' is denoted by $f_{mel}$ that corresponds to a frequency in Hz, which leads to the definition of MFCC. MFCC is a baseline acoustic feature set for speaker recognition applications [14], [26]. Delta-deltas and Deltas are the acceleration coefficients of MFCCs and derivatives, respectively. The following equation is applied to extract the Delta coefficients [27],

$$d_t = \frac{\sum_{n-1}^{N} n(c_{t+n} - c_{t-n})}{2 \sum_{n-1}^{N} n^2} \quad (2)$$

Given, the static Mel frequency cepstral coefficient of frame t that is denoted as $c_t$, and a Delta coefficient that is denoted by $d_t$. A standard value for N is 2. Acceleration coefficients (Delta-deltas) can be measured by employing the same equation, but they are calculated from the Deltas, not the MFCCs [27].

### B. Linear Predictive Coding (LPC)

Assume a frame of speech signal with N samples, $\{s_1, s_2, \ldots, s_n\}$. In LPC analysis, it is considered that the present sample is roughly predicted by a linear combination of p previous samples; i.e. [28],

$$\hat{s}_n = - \sum_{k=1}^{p} a_k \cdot s_{n-k} \quad (3)$$

where $\{a_1, \ldots, a_p\}$ are the LPC coefficients and p is the order of LPC analysis [28].

### C. Hybrid Algorithm DWPD

Wavelet transform has a powerful characteristic in using windows of different sizes. It accepts limited windows at high frequencies, whereas large windows are accepted at low frequencies. Therefore, this characteristic leads to the best time-frequency resolution in all frequency rates [29]. WPD and DWT utilize digital filtering methods to get a time-scale representation of the signals. DWT is characterized as follows [30],

$$w(j,k) = \sum_j \sum_k X(k) 2^{-\frac{j}{2}} \Psi(2^{-j} n - k) \quad (4)$$

where the basic analyzing function is denoted as Ψ(t), which is known as the mother wavelet.

WPD provides good frequency and time resolutions. As a result, it is effective in several disciplines of speech processing. WPD is a simplification of DWT and it is a more flexible and thorough technique than DWT. In WPD, the signal is divided into high-frequency components and low-frequency components at each level similar to DWT. However, the difference between WPD and DWT is that the DWT is employed to the low pass result only, whereas WPD uses the transform step to the high pass and the low pass result. The hybrid algorithm DWPD merges features of both WPD and DWT [30].

### D. Discrete Cosine Transform (DCT)

Discrete Cosine Transform can be applied for speech compression due to the high correlation in adjacent coefficients. A reconstruction of a sequence can be achieved precisely from very few DCT coefficients. This characteristic of DCT aids in the efficient reduction of data. The DCT of a 1-D sequence x(n) of length N is given as follows [31],

$$X(m) = [\frac{2}{N}]^{1/2} C_m \sum_{m=0}^{N-1} x(n) \cos[\frac{(2n+1)m\pi}{2N}] \quad (5)$$

where, m=0, 1, - - - - - -, N-1.

### E. Probabilistic Linear Discriminate Analysis (PLDA)

The scoring (accuracy) in PLDA is performed by applying the batch-likelihood ratio between a test i-vector (features of the model/speaker) and a target (the speaker identity) [32]. Given the i-vectors, $w_{target}$ and $w_{test}$, the batch likelihood ratio can be computed using the following equation [32],

$$\ln(\frac{P(w_{target}, w_{test}|H_1)}{P(w_{target}|H_0)P(w_{test}|H_0)}) \quad (6)$$

where $H_1$ signifies the assumption that the i-vectors exemplify the same speaker and $H_0$ signifies that they do not.

Table I summarizes the advantages and disadvantages of each feature extraction method.

TABLE I. ADVANTAGES AND DISADVANTAGES OF EACH OF FEATURES EXTRACTION TECHNIQUES

| Methods | Pros | Cons |
|---|---|---|
| MFCCs $\Delta^2$ [33] | Because the frequency ranges in MFCC are placed logarithmically, the human system response can be projected more precisely than any other system. | Because MFCCs tend to have low robustness in the presence of additive noise, the values in speech recognition models should be normalized to reduce the influence of noise. |
| LPC [33] | LPC can reduce the total of the squared deviations between the original and anticipated speech signals over a finite time period. | Because human perceptions differ in frequency, a speech recognition system based on LPC that approximates continuously weighted for the entire spectrum yields hidden results. |
| Hybrid Algorithm DWPD [30], [11] | Connects the features of both low-frequency and high-frequency components noted as DWT and WPD, respectively. DWPD can not only reduce the high-frequency range to more segments, but it can also avoid complications in the computation. | To get rid of the noise, the high frequencies are reduced in DWPD. However, the signal's high-frequency parts may contain useful information from time to time. |
| PLDA [34] | It is an adaptable acoustic technique that provides a wide range of noncorrelated input frames with no covariance modeling constraints. | The Gaussian supposition on the class conditional distributions is just that: a supposition. |
| DCT [35] | Removes redundancy from audio data to accelerate the model. | DCT uses only real-valued functions, making it inflexible. |

## III. Speech Database

Arabic Emirati-accented dataset is used in this work to evaluate the five feature extraction methods using Convolutional Neural Network-Support Vector Machine (CNN-SVM) classifier [36]. The Arabic corpus consists of 50 Emirati actors (25 males and 25 females) ranging in age from 20 to 55 years old who speak in an Emirati-accented dialect. Furthermore, the corpus retains five distinct emotions, including sadness, fear, happiness, disgust, and anger, in addition to the neutral condition. Each emotion is expressed in eight different ways. Nine times each utterance is repeated. In the United Arab Emirates, these expressions are frequently used. With the help of the "College of Communication at the University of Sharjah in the UAE," skilled engineers recorded the corpus. The corpus used in this study is displayed in Table II, with the Emirati accent statements in the left column and the equivalent English translation in the right column [37].

TABLE II. THE ARABIC EMIRATI CORPUS AND ITS ENGLISH INTERPRETATION

| Statement Number | "Emirati Sentences | English Translation |
|---|---|---|
| 1 | بنتلاقى وياك عقب ساعة | We will meet with you in an hour |
| 2 | سير عند أبويا يباك | Go to my father he wants you |
| 3 | هات تلفوني من الحجرة | Bring my cell phone from the room |
| 4 | مشغول/مشغولة الحين برمّسك عقب | I am busy now I will talk to you later |
| 5 | كل بيّاع يمدح سوقه | Every seller praises his market |
| 6 | الغريب ذيب وعضته ما تطيب | A stranger is a wolf whose bite wound will not heal |
| 7 | ناس احشمهم وناس احشم نفسك عنهم | Show respect around some people and show self-respect around other people |
| 8 | اللي ما قدرت تييبه لا تعيبه واللي ما تطوله لا تحوم حوله | Don't criticize what you can't get and don't swirl around something you can't obtain" |

Disguising the dataset was accomplished by Audacity® free software [38] on the 50 native Arabic Emirati-accented speakers. The fifty speakers were individually asked to utter the eight sentences. Each speaker was required to repeat each sentence nine times under six different emotions. The speakers were not permitted to exercise creating such utterances under any emotion. This database was recorded in a clean environment that was not influenced by any background noise.

The datasets contain both training and testing files. As a result, the training phase and the testing phase are the two key phases. Since our model is text-independent (texts in the training phase are not the same as in the testing), the training phase includes only the neutral state. The first 4 sentences are used in the training phase with 9 times repetition per sentence. The total size of our dataset is 12,600 sentences. The total number of used utterances in the training phase is 1,800 (50 speakers first four utterances 9 repetitions in neutral condition). The total number of used samples in the testing phase is 10,800 (50 speakers last four utterances repetitions 6 talking conditions)".

## IV. Results and Discussion

In this work, the selection of optimum features is accomplished through five well-known methods. These methods are MFCCs $\Delta^2$, hybrid algorithm DWPD, PLDA, LPC, and DCT. Experimental results are displayed in Tables III, IV, and V, for high-pitched, low-pitched, and EVC effects, respectively.

In Table III, MFCCs $\Delta^2$ shows 80.8% accuracy under the high-pitched effect, which is the highest accuracy amongst others. This was followed by LPC that reported 76.1% accuracy. Hybrid algorithm DWPD, PLDA, DCT are the three least techniques, where they showed 73.4%, 70%, and 65.3%, respectively.

In Table IV, the same methods are experienced under low-pitched effects. Likewise, MFCCs $\Delta^2$ remarked the highest performance (84.9%) in comparison with other feature extraction techniques. In contrast, the DCT method stated an accuracy of 69.8%. The low-pitched effect has witnessed the highest performance for all the methods.

The EVC effect has reported the lowest performances as shown in Table V. LPC and hybrid algorithm DWPD are the second-best accuracies after the MFCCs $\Delta^2$ method, where the achieved performances are 72.8% and 70.2%, respectively.

As a result, in the three disguising effects, MFCCs delta-delta performed the best among the other feature extraction approaches. The DCT approach, on the other hand, provides the least impressive results. LPC achieves the second-best approach, followed by the Hybrid Algorithm DWPD. Furthermore, the low-pitched effect reported the highest performance for all the methods amongst the high-pitched and EVC effects. This is followed by the high-pitched effect. EVC effect is the most effect that reduces the SI performance. In EVC effect, MFCCs $\Delta^2$ accuracy showed 76.3% compared to 84.9% in high-pitched effect, which is a difference of 8.6%.

Due to the transformation of audio into electrical signals for amplification, EVC remarked the lowest among the other effects. In contrast to EVC, the low and high-pitched effects only play with the original audio with altering the frequency wavelength.

TABLE III. THE PERFORMANCE OF FIVE DIFFERENT FEATURES EXTRACTION TECHNIQUES UNDER HIGH-PITCHED EFFECT

| Technique | Average Speaker Identification Accuracy |
|---|---|
| MFCCs $\Delta^2$ | 80.8% |
| LPC | 76.1% |
| Hybrid Algorithm DWPD | 73.4% |
| PLDA | 70.0% |
| DCT | 65.3% |

TABLE IV. THE PERFORMANCE OF FIVE DIFFERENT FEATURES EXTRACTION TECHNIQUES UNDER LOW-PITCHED EFFECT

| Technique | Average Speaker Identification Accuracy |
|---|---|
| MFCCs $\Delta^2$ | 84.9% |
| LPC | 80.0% |
| Hybrid Algorithm DWPD | 77.5% |
| PLDA | 75.4% |
| DCT | 69.8% |

TABLE V. THE PERFORMANCE OF FIVE DIFFERENT FEATURES EXTRACTION TECHNIQUES UNDER EVC EFFECT

| Technique | Average Speaker Identification Accuracy |
|---|---|
| MFCCs Δ2 | 76.3% |
| LPC | 72.8% |
| Hybrid Algorithm DWPD | 70.2% |
| PLDA | 67.3% |
| DCT | 63.9% |

## V. CONCLUSION

This paper presents five distinct feature extraction techniques, which are employed, explored, and analyzed thoroughly for speaker identification in disguised voices under emotional talking conditions. All evaluations are accomplished using Arabic Emirati-accented database. Three effects are employed in the evaluation, which are high-pitched, low-pitched, and Electronic Voice Conversion (EVC). Experimental results reported that the concatenated "MFCCs-delta-delta, MFCCs-delta, and MFCCs" is the best feature extraction method. On the other hand, DCT approach delivered the least impressive results. Additionally, low-pitched effect witnessed the highest performance for all the methods, whereas EVC experienced the lowest.

Limitations lay around the limited number of databases used, as well as the evaluation of further classifiers. Moreover, only five different feature extraction methods have been used. Our future work aims to use deep learning techniques to improve the research of disguised voices in the speaker recognition model under abnormal talking states in the future.